\documentclass[reprint, superscriptaddress]{revtex4-2}
\usepackage{graphicx}
\usepackage{braket}
\usepackage{amssymb,amsmath}
\usepackage[T1]{fontenc}
\usepackage[english]{babel}
\usepackage{textgreek} 
\usepackage{color}

\usepackage{natbib}
\begin{document}

\title{Strong-field quantum control in the extreme ultraviolet using pulse shaping}

\author{Fabian Richter}
\affiliation{Institute of Physics, University of Freiburg, Hermann-Herder-Str. 3, D-79104 Freiburg, Germany}
\author{Ulf Saalmann}
\affiliation{Max-Planck-Institut für Physik komplexer Systeme, N\"othnitzer Str. 38, 01187 Dresden,
Germany}
\author{Enrico Allaria}
\affiliation{Elettra-Sincrotrone Trieste S.C.p.A., 34149 Basovizza (Trieste), Italy}
\author{Matthias Wollenhaupt}
\affiliation{Institute of Physics, University of Oldenburg, Carl-von-Ossietzky-Str. 9-11, 26129
Oldenburg, Germany}
\author{Benedetto Ardini}
\affiliation{IFN-CNR, Dipartimento di Fisica, Piazza L. Da Vinci 32, 20133 Milan, Italy}
\author{Alexander Brynes}
\author{Carlo Callegari}
\affiliation{Elettra-Sincrotrone Trieste S.C.p.A., 34149 Basovizza (Trieste), Italy}
\author{Giulio Cerullo}
\affiliation{IFN-CNR, Dipartimento di Fisica, Piazza L. Da Vinci 32, 20133 Milan, Italy}
\author{Miltcho Danailov}
\author{Alexander Demidovich}
\affiliation{Elettra-Sincrotrone Trieste S.C.p.A., 34149 Basovizza (Trieste), Italy}
\author{Katrin Dulitz}
\affiliation{Institut für Ionenphysik und Angewandte Physik, Universit\"at Innsbruck, 6020 Innsbruck, Austria}
\author{Raimund Feifel}
\affiliation{Department of Physics, University of Gothenburg, Origov\"agen 6 B, SE-412 96 Gothenburg, Sweden}
\author{Michele Di Fraia}
\affiliation{Elettra-Sincrotrone Trieste S.C.p.A., 34149 Basovizza (Trieste), Italy}
\affiliation{Istituto Officina dei Materiali - CNR (CNR-IOM), Strada Statale 14 – km 163.5, Trieste, 34149, Italy}
\author{Sarang Dev Ganeshamandiram}
\affiliation{Institute of Physics, University of Freiburg, Hermann-Herder-Str. 3, D-79104 Freiburg, Germany}
\author{Luca Giannessi}
\affiliation{Elettra-Sincrotrone Trieste S.C.p.A., 34149 Basovizza (Trieste), Italy}
\affiliation{Istituto Nazionale di Fisica Nucleare - Laboratori Nazionali di Frascati, Via E. Fermi 40, 00044 Frascati, Roma}
\author{Nicolai G\"olz}
\author{Sebastian Hartweg}
\author{Bernd von Issendorff}
\affiliation{Institute of Physics, University of Freiburg, Hermann-Herder-Str. 3, D-79104 Freiburg, Germany}
\author{Tim Laarmann}
\affiliation{Deutsches Elektronen-Synchrotron DESY, Notkestr. 85, 22607
Hamburg, Germany}
\affiliation{The Hamburg Centre for Ultrafast Imaging CUI, Luruper Chaussee 149, 22761 Hamburg, Germany}
\author{Friedemann Landmesser}
\author{Yilin Li}
\affiliation{Institute of Physics, University of Freiburg, Hermann-Herder-Str. 3, D-79104 Freiburg, Germany}
\author{Michele Manfredda}
\affiliation{Elettra-Sincrotrone Trieste S.C.p.A., 34149 Basovizza (Trieste), Italy}
\author{Cristian Manzoni}
\affiliation{IFN-CNR, Piazza L. Da Vinci 32, 20133 Milan, Italy}
\author{Moritz Michelbach}
\author{Arne Morlok}
\affiliation{Institute of Physics, University of Freiburg, Hermann-Herder-Str. 3, D-79104 Freiburg, Germany}
\author{Marcel Mudrich}
\affiliation{Department of Physics and
Astronomy, Aarhus University, Ny Munkegade 120, DK-8000 Aarhus, Denmark}
\author{Aaron Ngai}
\affiliation{Institute of Physics, University of Freiburg, Hermann-Herder-Str. 3, D-79104 Freiburg, Germany}
\author{Ivaylo Nikolov}
\author{Nitish Pal}
\affiliation{Elettra-Sincrotrone Trieste S.C.p.A., 34149 Basovizza (Trieste), Italy}
\author{Fabian Pannek}
\affiliation{Institute for Experimental Physics, University of Hamburg, Luruper Chaussee 149, 22761
Hamburg, Germany}
\author{Giuseppe Penco}
\author{Oksana Plekan}
\author{Kevin C. Prince}
\affiliation{Elettra-Sincrotrone Trieste S.C.p.A., 34149 Basovizza (Trieste), Italy}
\author{Giuseppe Sansone}
\affiliation{Institute of Physics, University of Freiburg, Hermann-Herder-Str. 3, D-79104 Freiburg, Germany}
\author{Alberto Simoncig}
\affiliation{Elettra-Sincrotrone Trieste S.C.p.A., 34149 Basovizza (Trieste), Italy}
\author{Frank Stienkemeier}
\affiliation{Institute of Physics, University of Freiburg, Hermann-Herder-Str. 3, D-79104 Freiburg, Germany}
\author{Richard James Squibb}
\affiliation{Department of Physics, University of Gothenburg, Origov\"agen 6 B, SE-412 96 Gothenburg, Sweden}
\author{Peter Susnjar}
\author{Mauro Trovo}
\affiliation{Elettra-Sincrotrone Trieste S.C.p.A., 34149 Basovizza (Trieste), Italy}
\author{Daniel Uhl}
\affiliation{Institute of Physics, University of Freiburg, Hermann-Herder-Str. 3, D-79104 Freiburg, Germany}
\author{Brendan Wouterlood}
\affiliation{Institute of Physics, University of Freiburg, Hermann-Herder-Str. 3, D-79104 Freiburg, Germany}
\author{Marco Zangrando}
\affiliation{Elettra-Sincrotrone Trieste S.C.p.A., 34149 Basovizza (Trieste), Italy}
\affiliation{Istituto Officina dei Materiali - CNR (CNR-IOM), Strada Statale 14 – km 163.5, Trieste, 34149, Italy}
\author{Lukas Bruder}
\email{lukas.bruder@physik.uni-freiburg.de}
\affiliation{Institute of Physics, University of Freiburg, Hermann-Herder-Str. 3, D-79104 Freiburg, Germany}

\date{\today}

\begin{abstract}
Tailored light-matter interactions in the strong coupling regime enable the manipulation and control of quantum systems with up to unit efficiency\,\cite{bergmann_coherent_1998, vitanov_laser-induced_2001}, with applications ranging from quantum information to photochemistry\,\cite{saffman_quantum_2010, becker_ultrafast_2016, pedrozo-penafiel_entanglement_2020, feist_quantum_2015, glaser_training_2015}.
While strong light-matter interactions are readily induced at the valence electron level using long-wavelength radiation\,\cite{brabec_intense_2000}, comparable phenomena have been only recently observed with short wavelengths, accessing highly-excited multi-electron and inner-shell electron states\,\cite{ott_strong-field_2019, kanter_unveiling_2011}. 
However, the quantum control of strong-field processes at short wavelengths has not been possible, so far, due to the lack of pulse shaping technologies\,\cite{weiner_femtosecond_2000} in the extreme ultraviolet (XUV) and X-ray domain. 
Here, exploiting pulse shaping of the seeded free-electron laser (FEL) FERMI, we demonstrate the strong-field quantum control of ultrafast Rabi dynamics in helium atoms with high fidelity. 
Our approach unravels a strong dressing of the ionization continuum, otherwise elusive to experimental observables. 
The latter is exploited to achieve control of the total ionization rate, with prospective applications in many XUV and soft X-ray experiments.
Leveraging recent advances in intense few-femtosecond to attosecond XUV to soft X-ray light sources\,\cite{pellegrini_physics_2016, orfanos_non-linear_2020}, our results open an avenue to the efficient manipulation and selective control of core electron processes and electron correlation phenomena in real time.
\end{abstract}

\maketitle


\section{Introduction}
Strong-field phenomena play an important role in our understanding of the quantum world. 
Light-matter interactions beyond the perturbative limit can substantially distort the energy landscape of a quantum system, which forms the basis of many intriguing strong-field effects\,\cite{brabec_intense_2000} and opens an avenue for efficient quantum control schemes\,\cite{sussman_dynamic_2006}. 
Moreover, resonant strong coupling induces rapid Rabi cycling of the level populations\,\cite{rabi_space_1937}, enabling complete population transfer to a target state\,\cite{vitanov_laser-induced_2001}.  
The development of intense XUV and X-ray light sources\,\cite{pellegrini_physics_2016, orfanos_non-linear_2020} has recently permitted the investigation of related phenomena beyond valence electron dynamics, in highly excited, multi-electron and inner-shell electron states\,\cite{young_femtosecond_2010, kanter_unveiling_2011, ott_reconstruction_2014, glover_controlling_2010, ranitovic_controlling_2011,fushitani_femtosecond_2016, ott_strong-field_2019}. 
Yet in most of these studies, the dressing of the quantum systems was induced by intense infrared fields overlapping with the XUV/X-ray pulses. 
In contrast, the alteration of energy levels directly by short-wavelength radiation is more difficult. 
So far, only a few studies reported XUV-induced AC-Stark shifts of moderate magnitude ($\lesssim 100$\,meV), difficult to resolve experimentally\,\cite{meyer_two-photon_2010, sako_suppression_2011, flogel_rabi_2017, ott_strong-field_2019, ding_nonlinear_2019, nandi_observation_2022}. 

The active control of quantum dynamics using tailored light fields marks another important quest in exploring and mastering the quantum world\,\cite{stuart_a_rice_optical_2000, tannor_introduction_2007, shapiro_quantum_2012}. 
At long wavelengths, sophisticated pulse shaping techniques facilitate the precise quantum control and even the adaptive-feedback control of many light-induced processes, both in the weak- and strong-field regime\,\cite{brif_control_2010, brixner_quantum_2003, dantus_experimental_2004,lozovoy_systematic_2005, wollenhaupt_femtosecond_2005, silberberg_quantum_2009}. 
Several theoretical studies have pointed out the potential of pulse shaping in XUV and X-ray experiments\,\,\cite{greenman_laser_2015,goetz_quantum_2016,keefer_selective_2021}. 
As a first experimental step in this direction, phase-locked mono- and poly-chromatic pulse sequences have been generated\,\cite{prince_coherent_2016,usenko_attosecond_2017, usenko_attosecond_2017, wituschek_tracking_2020,heeg_coherent_2021}. 
Using this tool, first coherent control demonstrations in the perturbative limit\,\cite{prince_coherent_2016,koll_experimental_2022,heeg_coherent_2021} and the generation of intense attosecond pulses were achieved\,\cite{maroju_attosecond_2020, maroju_attosecond_2023-1}. 
Moreover, ultrafast polarization shaping at XUV wavelengths\,\cite{perosa_femtosecond_2023} 
and chirp control for the temporal compression of XUV pulses\,\cite{gauthier_spectrotemporal_2015} were recently demonstrated.
However, spectral phase shaping, which forms the core of pulse shaping techniques, has not been demonstrated for the control of quantum phenomena at short wavelengths. 
Here, we establish spectral phase shaping of intense XUV laser pulses and demonstrate high fidelity quantum control of the Rabi and photoionization dynamics in helium. 

\section{Results}
\begin{figure}
\centering\includegraphics[width=\linewidth]{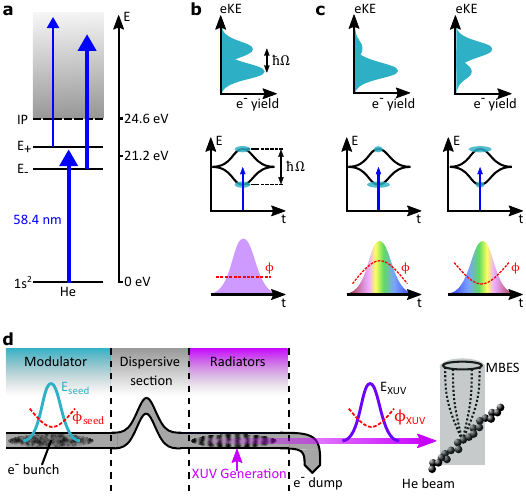}
\caption{XUV strong-field coherent control scheme. (a) Intense XUV pulses dress the He $1s^2$, $1s2p$ states and the electron continuum. 
$E_\pm$ labels indicate the bound dressed states correlated to the $1s2p$ bare state. 
Mixing of p- and d-waves in the dressed continuum results in different coupling strengths to the dressed bound states (indicated by thickness of arrows). 
(b,c)  
In the time-domain the AT splitting follows the intensity profile of the XUV field (middle panel).  
The dressed-state populations are monitored in the photoelectron eKE distributions (upper panel). 
XUV pulse shaping enables the control of the non-perturbative quantum dynamics (lower panel). For a flat phase $\phi$ (no chirp), both excited dressed states are equally populated. For a positive phase curvature (up chirp) the population is predominantly transferred to the lower dressed state and the upper state is depleted, while for negative curvature (down chirp) the situation is reversed. 
(d) Principle of XUV pulse shaping at the FEL FERMI. 
Intense seed laser pulses overlap spatially and temporally with the relativistic electron bunch in the modulator section of the FEL, leading to a modulation in the electron phase space. 
The induced energy modulations are converted into electron-density oscillations upon passing a dispersive magnet section. 
The micro-bunched electrons then propagate through a section of radiator undulators, producing a coherent XUV pulse. 
In this process the phase function of the seed pulse is coherently transferred to the XUV pulse, resulting in precise XUV phase shaping. 
The FEL pulses are focused into the interaction volume, exciting and ionizing He atoms. 
The photoelectrons are detected with a magnetic bottle electron spectrometer (MBES). 
}
\label{fig1}
\end{figure}
In the experiment, He atoms are dressed and ionized by intense coherent XUV pulses ($I > 10^{14}$\,W/cm$^2$) delivered by the seeded FEL FERMI (Fig.\,\ref{fig1}a). 
The high radiation intensity also causes a strong dressing of the photoelectron continuum, 
while the ionization dynamics of the atomic system are still in the multiphoton regime (Keldysh parameter $\gamma = 11$). 
In contrast, the dynamics of a system dressed with NIR radiation of comparable intensity, would be dominated by tunnel and above barrier ionization ($\gamma = 0.35$)\,\cite{brabec_intense_2000}. 
Hence, the use of short-wavelength radiation provides access to a unique regime, where the interplay between strongly dressed bound states and a strongly dressed continuum can be studied.

To dress the He atoms, we induce rapid Rabi cycling of the $1s^2 \rightarrow 1s2p$ atomic resonance with a near-resonant field $E(t)$. 
The generalized Rabi frequency of this process is $\Omega = \hbar^{-1} \sqrt{(\mu E)^2+\delta^2}$, where $\mu$ denotes the transition dipole moment of the atomic resonance and $\delta$ the energy detuning.  
In the dressed-state formalism\,\cite{cohen-tannoudji_atom-photon_1998} the eigenenergies of the bound states depend on the field intensity and show the characteristic Autler-Townes (AT) energy splitting $\Delta E = \hbar \Omega$\,\cite{autler_stark_1955}.  
The observation of this phenomenon requires the mapping of the transiently dressed level structure of He while perturbed by the external field\,\cite{wollenhaupt_femtosecond_2006}. 
This is achieved by immediate photoionization over the course of the femtosecond pulses, thus projecting the time-integrated energy level shifts onto the electron kinetic energy (eKE) distribution (Fig.\,\ref{fig1}b).

In analogy to the bound state description, the dressed continuum states are obtained by a diagonalization of the corresponding Hamiltonian. 
The hybrid electron-photon eigenstates compose of a mixing of partial waves with different angular momenta, which alters the coupling strength to the dressed bound states of the He atoms (Fig.\,\ref{fig1}a).

\begin{figure}
\centering\includegraphics[width=\linewidth]{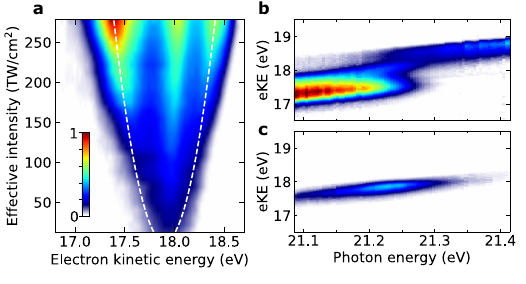}
\caption{Build-up of the Autler-Townes splitting in He atoms. 
(a) Detected photoelectron eKE distribution (raw data) as a function of the XUV intensity (FEL photon energy: 21.26\,eV, GDD = 135 fs$^2$). 
Dashed lines show the calculated AT splitting for an effective XUV peak intensity $I_\mathrm{eff}$ accounting for the spatial averaging in the interaction volume (values given on y-axis). 
(b) Photoelectron spectra as a function of photon energy recorded for high XUV intensity ($I_\mathrm{eff}=2.92(18)\times10^{14}$\,W/cm$^2$) and in (c) for lower intensity ($I_\mathrm{eff} \approx 10^{13}$\,W/cm$^2$). 
In (b), a clear avoided crossing between the lower/higher AT band is visible directly in the raw photoelectron spectra. 
The photoelectron distribution peaking at eKE=\,17.9\,eV in (a) and (b) is ascribed to He atoms excited by lower XUV intensity (see text).
}
\label{fig2}
\end{figure}
Figure\,\ref{fig2} demonstrates experimentally the dressing of the He atoms. 
The build-up of the AT doublet is clearly visible in the raw photoelectron spectra as the XUV intensity increases (Fig.\,\ref{fig2}a). 
The evolution of the AT doublet splitting is in good agreement with the expected square-root dependence on the XUV intensity $\Delta E = \mu \sqrt{2I_\mathrm{eff}/(\epsilon_0c)}$. 
Here, $I_\mathrm{eff}$ denotes an effective peak intensity, accounting for the spatially averaged intensity distribution in the interaction volume, $\epsilon_0, \, c$ denote the vacuum permittivity and the speed of light, respectively. 
The data can be thus used for gauging the XUV intensity in the interaction volume, a parameter otherwise difficult to determine. 
At the maximum XUV intensity, the photoelectron spectrum shows an energy splitting exceeding 1\,eV, indicative of substantial AC-Stark shifts in the atomic level structure. 
The large AT splitting further implies that a Rabi flopping within 2\,fs is achieved, offering a perspective for rapid population transfer outpacing possible competing intra and inter atomic decay mechanisms, which are ubiquitous in XUV and X-ray applications\,\cite{jahnke_interatomic_2020}.

\begin{figure*}
\centering\includegraphics[width=0.9\textwidth]{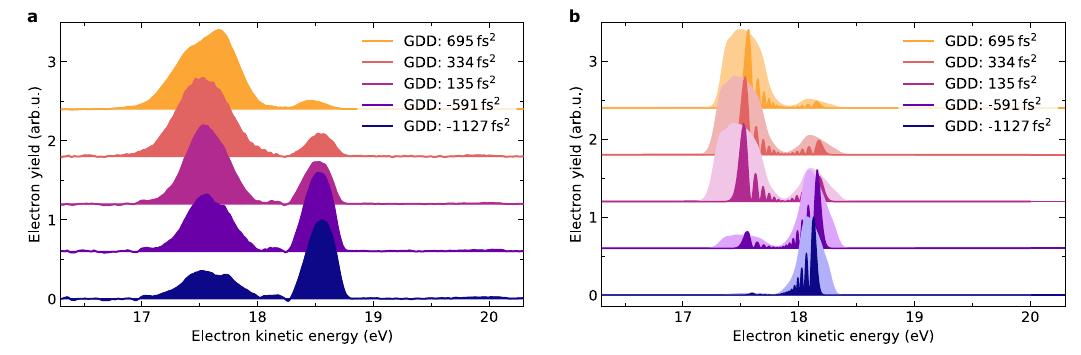}
\caption{Strong-field quantum control of dressed He populations for the He $1s^2 \rightarrow 1s2p$ resonance (photon energy: 21.25\,eV, $I_\mathrm{eff} = 2.8(2) \times 10^{14}$W/cm$^2$). 
(a) Photoelectron spectra obtained for phase-shaped XUV pulses (see given GDD values). 
The control of the dressed state populations is directly reflected in the relative change of amplitude in the photoelectron bands. 
The small peak at 18.13\,eV is due to imperfect removal of the lower intensity contribution from the aberrated focus. 
(b) TDSE-SAE calculations for a single laser intensity corresponding to the experimental $I_\mathrm{eff} = 2.8 \times 10^{14}$W/cm$^2$ (dark colors). 
Spectral fringes reflect here the temporal progression of the Rabi frequency during the light-matter interaction. 
The broadened photoelectron spectra (light colors) account for experimental broadening effects caused by the focal intensity averaging and the instrument response function.
The model underestimates the atomic dipole moment which leads to a factor of 1.4 smaller AT energy splitting compared to the experimental data.}
\label{fig3}
\end{figure*}

\begin{figure}
\centering\includegraphics[width=\linewidth]{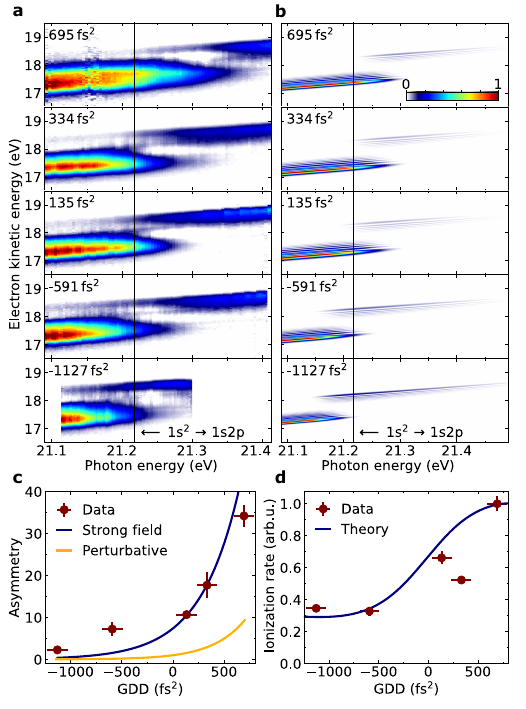}
\caption{Energy-domain representation of the quantum control scheme. 
(a) Photoelectron spectra as a function of energy detuning for different GDD values as labeled ($I_\mathrm{eff}=2.92(18)\times10^{14}$\,W/cm$^2$). 
(b) TDSE-SAE calculations. Broadening by the instrument response function is omitted in the model.  
(c) Amplitude ratio between upper/lower photoelectron bands evaluated at the $1s^2 \rightarrow 1s2p$ resonance, hence $\delta = 0$. 
Experimental data (red), TDSE-SAE model treating the bound and continuum dynamics non-perturbatively (blue), TDSE-SAE model applied to the bound state dynamics, but treating the continuum perturbatively (yellow). 
(d) Dependence of the He ionization rate on the spectral phase of the driving field. Data (red), TDSE-SAE model (blue). 
}
\label{fig4}
\end{figure}

Figures\,\ref{fig2}b,c show the photoelectron yield as a function of excitation photon energy. 
For high XUV intensity (Fig.\,\ref{fig2}b), the photoelectron spectra reveal an avoided level crossing of the dressed He states as they are mapped to the electron continuum (see also Fig.\,\ref{fig4}).  
Accordingly, at lower XUV intensity (Fig.\,\ref{fig2}c), the avoided crossing is not visible anymore. 
In the latter, the eKE distribution centers at $17.9$\,eV. 
In Fig.\,\ref{fig2}b, a similar contribution appears at the same kinetic energy which overlays the photoelectrons emitted from the strongly dressed atoms. 
Likewise, a significant portion of photoelectrons at eKE$\,\approx 17.9$\,eV in Fig.\,\ref{fig2}a does not show a discernible AT splitting. 
We conclude that a fraction of He atoms in the ionization volume are excited by much lower FEL intensity, which is consistent with the abberated intensity profile of the FEL measured in the ionization volume (Supp. Info. I). 
This overlapping lower intensity contribution does not influence the interpretation of the results in this work. 
For better visibility of the main features, we thus subtract this contribution from the data shown in Fig.\,\ref{fig3} and \ref{fig4}. 

The demonstrated dressing of He atoms provides the prerequisite for implementing the strong-field quantum control scheme, illustrated in Fig.\,\ref{fig1}b,c.  
The main mechanism underlying the control scheme is described in the framework of the selective population of dressed states (SPODS), which is well established in the NIR spectral domain\,\cite{Bayer_ch6}. 
Here, we extend SPODS to the XUV domain and include a new physical aspect, that is the transition of the bound atomic system into a strongly-dressed continuum. 
In SPODS, a flat phase leads to an equal population of both dressed states in the excited state manifold of helium, whereas a positive/negative phase curvature results in a predominant population of the lower/upper dressed state, respectively (Fig.\,\ref{fig1}c). 
The scheme has been experimentally demonstrated with long-wavelength radiation\,\cite{wollenhaupt_quantum_2006-1}, where pulse shaping techniques are readily available. 
However, the opportunities of pulse shaping technologies are largely unexplored for XUV and X-ray radiation. 

We solve this problem by exploiting the potential of seeded FELs to allow for the accurate control of XUV pulse properties\,\cite{gauthier_spectrotemporal_2015,de_ninno_single-shot_2015, gauthier_chirped_2016}. 
These demonstrations have been so far limited to applications of temporal compression and amplification of the FEL pulses. 
In contrast, the deterministic control of quantum dynamics in a material system involves many more degrees of freedom, which makes the situation considerably more complex.  
The seeded FEL FERMI operation is based on the high-gain harmonic generation (HGHG) principle\,\cite{allaria_highly_2012}, where the phase of an intense seed laser pulse is imprinted into a relativistic electron pulse to precondition the coherent XUV emission at harmonics of the seed laser (Fig.\,\ref{fig1}d). 
For FEL operation in the linear amplification regime, the phase $\phi_{nH}(t)$ of the FEL pulses emitted at the $n$'th harmonic of the seed laser follows the relationship\,\cite{gauthier_spectrotemporal_2015} 
\begin{equation}
    \phi_{nH}(t)\approx n[\phi_s(t)+\phi_e(t)] + \phi_a\, .
\end{equation} 
Here, $\phi_s$ denotes the phase of the seed laser pulses, which can be tuned with standard pulse shaping technology at long wavelengths (see Methods for details). 
$\phi_e$ accounts for the possible phase shifts caused by the energy dispersion of the electron beam through the dispersive magnet and is negligible for the parameters used in the experiment. 
$\phi_a$ accounts for the FEL phase distortion due to the amplification and saturation in the radiator and has been kept negligibly small by properly tuning the FEL (see Methods for details). 
While complex phase shapes may be implemented with this scheme, for the current objective of controlling the strong-field induced dynamics in He atoms, shaping the quadratic phase term (group delay dispersion - GDD) is sufficient\,\cite{wollenhaupt_quantum_2006-1}. 
Therefore, we focus on the GDD-control in the following discussion. 

Figure\,\ref{fig3} demonstrates the quantum control of the dressed He populations. 
The eKE distribution shows a pronounced dependence on the GDD of the XUV pulses (Fig.\,\ref{fig3}a). 
At minimum chirp (GDD= 135\,fs$^2$), we observe an almost even amplitude in the AT doublet, whereas for GDD $< 0$ the higher energy photoelectron band dominates and for GDD $> 0$ the situation is reversed. 
These changes directly reflect the control of the relative populations in the upper/lower dressed state of the He atoms. 
We obtain an excellent control contrast and the results are highly robust (Supp. Info. II), which is remarkable given the complex experimental setup.

The experiment is in good agreement with the theoretical model (Fig.\,\ref{fig3}b) numerically solving the time-dependent Schr\"odinger equation for a single active electron (TDSE-SAE, see Methods for details). 
To account for experimental broadening effects, we calculated the photoelectron spectra for a single intensity (corresponding to the experimental $I_\mathrm{eff}$) and including the focal intensity average present in the experiment (see Methods for details). 
All salient features of the experiment are well reproduced: 
The control of the dressed state populations is in very good qualitative agreement. 
The different widths and shapes of the photoelectron peaks are qualitatively well matched between experiment and the calculations. 
The difference in the AT energy splitting between experiment ($\Delta E_\mathrm{exp}\approx 1.02$\,eV) and theory ($\Delta E_\mathrm{theo}=0.74$\,eV) is in good agreement with the fact, that the model underestimates the transition dipole moment of the $1s^2 \rightarrow 1s2p$ transition by a factor of 1.4 (see Methods section). 

The high reproducibility, the excellent control contrast and the good agreement with theory confirm the feasibility of precise pulse shaping in the XUV domain and of quantum control applications, even of transient strong-field phenomena. 
This marks an important achievement in view of quantum optimal control applications at short wavelengths. 

We note, that the implemented control scheme is not restricted to adiabatic processes\,\cite{Bayer_ch6}. 
In fact, in the presented experiment the dynamics are only adiabatic for the largest frequency chirp (which corresponds to GDD = -1127\,fs$^2$) (see Supp. Info. III). 
However, this also shows that the condition for rapid adiabatic passage\,\cite{vitanov_laser-induced_2001} can be generally reached with our approach, unlocking new possibilities for efficient population transfer in the XUV and potentially in the soft X-ray regime. 

The active control of quantum dynamics with tailored light fields is one asset of pulse shaping. 
As another asset, systematic studies with shaped laser pulses can be used to unravel underlying physical mechanisms otherwise hidden\,\cite{bardeen_selective_1995}. 
Here, we demonstrate this concept for pulse shaping in the XUV domain. 
The high XUV intensities used in our study lead to a peculiar scenario in which both bound and continuum states are dressed and a complex interplay between their dynamics arises. 
Hence, for a comprehensive understanding of the strong-field physics taking place, the bound-state dynamics and the non-perturbative photoionization have to be considered. 
This is in contrast to the strong-field control at long wavelengths, where the continuum could be described perturbatively\,\cite{wollenhaupt_quantum_2006-1}. 

Fig.\,\ref{fig4}a,b show the avoided crossing of the photoelectron bands for different spectral phase curvatures applied to the XUV pulses. 
The experimental data reveals a clear dependence of the AT doublet amplitudes on the detuning and the GDD of the driving field, in good agreement with theory. 
In the strong dressing regime, the bound-continuum coupling marks a third factor which influences the photoelectron spectrum. 
As predicted by theory, the strong-field induced mixing of continuum states (Fig.\,\ref{fig1}a) leads to different photoionization probabilities for the upper/lower dressed state of the bound system\,\cite{saalmann_adiabatic_2018}. 
This is in agreement with the prevalent asymmetry of the AT doublet amplitudes observed in our data and calculations (Fig.\,\ref{fig4}a,b). 
An analogous effect is observed for the strong-field bound-continuum coupling in solid state systems\,\cite{mahmood_selective_2016}.  

To disentangle this strong-field effect from the influence of the detuning and spectral phase of the driving field, we evaluate the amplitude ratio between the upper/lower photoelectron bands at detuning $\delta = 0$\,eV (Fig.\,\ref{fig4}c). 
Interpolation to GDD $=0$\,fs$^2$ isolates the asymmetry solely caused by the strong-field bound-continuum coupling. 
We find reasonable agreement with our model when including the dressing of the ionization continuum (blue curve), in stark contrast to the same model but treating the continuum perturbatively (yellow curve). 
Hence, the dressing of the He atoms provides here a probe of the strong-field dynamics in the continuum. 
This property is otherwise difficult to access and becomes available by our systematic study of the spectral phase dependence on the photoelectron spectrum. 

Another possible mechanism for a general asymmetry in the AT doublet amplitudes could be the interference between ionization pathways via resonant and near-resonant bound states as recently suggested for the dressing of He atoms with XUV\,\cite{nandi_observation_2022, olofsson_photoelectron_2023-1} and for alkali atoms with bichromatic near infrared fields\,\cite{bayer_bichromatic_2024-1}. 
In our experiment , we study the energetically well-isolated transition $1s^2 \rightarrow 1s2p$, where contributions from neighboring optically active states should be negligible. 
This provides us a clean two-level system and greatly simplifies the data interpretation. 
For confirmation, we performed a calculation with a modified model in which any two-photon ionization via near-resonant states (except for the $1s2p$ state) was suppressed, and, thus, possible photoionization interference effects are eliminated. 
Still, we observe a pronounced asymmetry in the AT doublet amplitudes (see Supp. Info. IV). 
Moreover, due to the large Keldysh parameter ($\gamma = 11$) and the low ponderomotive potential ($U_p < 100\,$meV) in our study, other strong-field effects are expected to play a negligible role in the observed dynamics.
We thus assign the experimental observation to the coupling of the dressed atom dynamics with a dressed ionization continuum induced by intense XUV driving fields. 

A comprehensive understanding of the strong-field induced dynamics in the system lays the basis for another quantum control effect, that is the suppression of the system's ionization rate, as proposed theoretically\,\cite{saalmann_adiabatic_2018}. 
The excitation probability for one-photon transitions is generally independent of the chirp direction of the driving field. 
However, if driving a quantum system in the strong-field limit, its quasi-resonant two-photon ionization rate may become sensitive to the chirp direction. 
We demonstrate the effect experimentally in Fig.\,\ref{fig4}d. 
A substantial reduction of the He ionization rate by 64\,\% is achieved, solely by tuning the chirp of the FEL pulses while keeping the pulse area constant. 
The good agreement with the TDSE-SAE calculations confirms the mechanism. 
This control scheme exploits the interplay between the bound-state dynamics and the above discussed selective coupling of the upper/lower dressed state to the ionization continuum. 
We note a stabilization mechanism of the dressed states in He was recently proposed, effectively causing also a suppression of the ionization rate\,\cite{olofsson_photoelectron_2023-1}. 
This mechanism requires, however, extreme pulse parameters, difficult to achieve experimentally. 
In contrast, our approach based on shaped pulses is more feasible and applies to a broader parameter range. 

With this work, we have established a new tool for the manipulation and control of matter using XUV light sources. 
The demonstrated concept offers a wide pulse shaping window w.r.t. pulse duration, photon energy and more complex phase shapes. 
In particular, the recent progress in echo-enabled harmonic generation\,\cite{rebernik_ribic_coherent_2019, mirian_spectrotemporal_2020} promises to extend the pulse shaping concept to the soft X-ray domain ($\approx$ 600\,eV) where localized core-electron states can be addressed. 
As such, we expect our work will stimulate other experimental and theoretical activities exploring the exciting possibilities offered by XUV and soft X-ray pulse shaping: first theory proposals in this direction have already been made\,\cite{greenman_laser_2015,goetz_quantum_2016,keefer_selective_2021}. 
The demonstrated scheme already sets the basis for highly efficient adiabatic population transfer\,\cite{bergmann_coherent_1998,vitanov_laser-induced_2001} and an extension to cubic or sinusoidal phase shaping would open-up many more interesting control schemes\,\cite{lozovoy_systematic_2005, silberberg_quantum_2009}. 
This may find applications, e.g. in valence-core stimulated Raman scattering\,\cite{oneal_electronic_2020} or in efficient and fast qubit manipulation with XUV and soft X-ray light. 
Besides, selective control schemes may reduce the influence of competing ionization processes ubiquitous in XUV/X-ray spectroscopy and imaging experiments, where our work provides the first experimental demonstration in this direction. 
The generation of coherent attosecond pulse trains, with independent control of amplitude and phases, has been demonstrated at seeded FELs\,\cite{maroju_attosecond_2020}, bringing pulse shaping applications on the attosecond time scale within reach. 
This opens an exciting avenue towards the quantum control of molecular and solid state systems with chemical selectivity and on attosecond time scales.

\section{Methods}
\subsection{Experiment}
The experiments were performed at the low density matter (LDM) endstation\,\cite{lyamayev_modular_2013} of the FEL FERMI-1\,\cite{allaria_highly_2012}.
The FEL was operated in circular polarization at the 6th harmonic of the seed laser. 
The FEL photon energy was tuned in the 21.05\,eV to 21.47\,eV range with an optical parametric amplifier in the seed laser setup. 
Maximum pulse energy at the target was $E_\mathrm{max}= 71$\,\textmu J, taking transmission losses into account. 
A N$_2$-gas filter was used for continuous attenuation of the pulse energy. 
For the data in Fig.\,\ref{fig2}c, a Sn-filter (thickness: 200\,nm) was inserted, attenuating the XUV intensity by roughly one order of magnitude.  
At minimum chirp setting (GDD\,$= 135\,$fs$^2$), an FEL pulse duration of 49(3)\,fs was measured by a cross-correlation between the FEL pulses and an 800-nm auxiliary pulse. 
The beam size at the target was 8.00(8) $\times$ 11.3(1)\,\textmu m$^2$, reconstructed with a Hartmann wavefront sensor. 
Assuming a Gaussian spatial mode, this yields a calculated estimate for the maximum reachable peak intensity of 3.84$\times$10$^{14}$\,W/cm$^2$ at the interaction region. 
In comparison, the effective intensity deduced from the AT splitting is $I_\mathrm{eff}=2.78(2)\times10^{14}$\,W/cm$^2$. 
This value is 27\,\% smaller than the value calculated for a Gaussian spatial mode, hinting at an abberated spatial mode (see also Supp. Info. I). 

Spectral phase shaping of the seed laser was implemented by tuning a single-pass transmission grating compressor and characterized by self-diffraction frequency-resolved optical gating (SD-FROG). 
In the applied tuning range, changes of higher-order phase terms are small and are thus neglected. 
The coherent transfer of the seed phase $\phi_s$ to the FEL phase $\phi_{nH}$ was with FEL simulations using the GENESIS 1.3 code\,\cite{reiche_genesis_1999} and a FEL model\,\cite{pannek_accurate_2023}.  
With these tools, the FEL was analyzed as outlined in Ref.\,\cite{gauthier_spectrotemporal_2015} for a set of seed laser and FEL settings prior to the beamtime. 
Details can be found in Ref.\,\cite{pannek_accurate_2023}. 
To minimize the additional chirp introduced by the FEL amplification process ($\phi_a$), the FEL amplification was kept reasonably low and only five (out of six) undulators were used. 
At these conditions, $\phi_a$ is supposed to be negligible. 
With these precautions the major source of uncertainty in the GDD comes from the exact setting of the FEL and seeding parameters. 
According to our simulations we can estimate the uncertainty on the GDD of the FEL to be $\pm 100$\,fs$^2$.  

At the endstation, a pulsed valve was used at room temperature to create a pulsed beam of He atoms synchronized with the arrival of the XUV pulses. 
In the interaction region, the atomic beam intersected the laser pulses perpendicularly and the generated photoelectrons were detected with a magnetic bottle electron spectrometer (MBES). 
A retardation potential of 14\,eV was applied to optimize the detector resolution. 
For the FEL settings used, the contribution of second harmonic FEL emission to the ionization yield is expected to be at least three orders of magnitude smaller and can thus be neglected.
For the experimental parameters used, space charge effects can be neglected as confirmed by measurements with different atom densities in the ionization volume. 
A distortion of photoelectron trajectories by the large retardation potentials was ruled out by simulations of the electron trajectories.

\subsection{Theory}
In order to calculate the photoelectron spectra we solve the time-dependent Schr\"odinger equation (TDSE) for a single-active-electron (SAE) model of the He atom. The effective potential in this model reads
\begin{equation}
V(r)= -\frac {1}{r}\big[1 + \mathrm{e}^{-r/r_{0}} - r\mathrm{e}^{-r/r_{1}}\big]\, ,
\end{equation}
where $r$ denotes the radial coordinate. 
It has the correct asymptotic behavior for $r\,{\to}\,0$ and $r\,{\to}\,\infty$ and
the values of $r_{0}\,{=}\,0.5670$\,\AA\ and $r_{1}\,{=}\,0.4396$\,\AA\ guarantee that the binding energies $E_{1\mathrm{s}^{2}}\,{=}\,{-}24.5874$\,eV and $E_{1\mathrm{s}2\mathrm{p}}\,{=}\,{-}3.3694$\,eV of He \cite{kramida_a._nist_2015} are reproduced.
The dipole moment in this model is a factor of 1.4 smaller than the NIST value \cite{kramida_a._nist_2015}, which is the reason for the smaller AT splitting obtained in calculations compared to the experimental data. 
Field-free eigenstates up to an angular momentum of $\ell{=}3$ are calculated in a box of radius $R\,{=}\,1.69{\times}10^{4}$\,\AA\ by means of the Numerov method and are used as a basis for the TDSE, which is solved in the velocity form. 
The box size $R$ is chosen sufficiently large to omit the need of absorbing boundary conditions.
Thus, photoelectron spectra can be calculated directly from the occupations of the field-free eigenstates obtained in the propagation. 
Due to the high intensities of interest we treat the vector potential 
of the FEL pulse classically and use a Gaussian envelope. 
 
Thus the vector potential reads 
\begin{subequations}\label{eq:pulse}\begin{align} 
\mathbf{A}(t)&=A_{0}\,g(t)\,\big\{\Re f(t),\Im f(t),0\big\} 
\tag{\ref{eq:pulse}} 
\\ 
g(t) &= \exp\big({-}2\ln2\, t^2/T^2\big) 
\\ 
f(t) &= \exp\big(i(\omega_0 t+a t^2)\big), 
\end{align}\end{subequations} 
where $T$ is the FEL pulse duration, which depends on the chirp, 
$\omega_{0}$ denotes the carrier frequency and 
$a$ is the linear chirp rate, that relates to the quadratic 
spectral phase coefficient $\phi_{2}$ 
(i.e.\ the group delay dispersion\,--\,GDD) as 
\begin{equation} 
a = \frac{\phi_{2}}{2\phi_{2}{\!}^{2}+(T_{0}{\!}^{2}/\sqrt{8}\ln2)^{2}}. 
\end{equation} 
Here $T_{0}$ denotes the Fourier-transform-limited pulse duration.

To account for the experimental response function and the focal intensity averaging in the experiment, we calculated the average of the photoelectron spectra for a range of laser intensities ($8.3\times 10^{13} - 6.9 \times 10^{14}$\,W/cm$^2$) and convoluted the result by the instrument response function ($\approx 50\,$meV). 
In this way, the average intensity in the calculations is $2.74 \times 10^{14}$\,W/cm$^2$, which matches the effective intensity in the experiment of $I_\mathrm{eff} = 2.8 \times 10^{14}$\,W/cm$^2$. 
These computationally intense simulations were performed for a few laser wavelengths and were used to calculate the data in Fig.\,\ref{fig3}b, \ref{fig4}c,d. 
We omitted a calculation of all laser wavelengths disclosed in Fig.\,\ref{fig4}b. 
Here, we show the calculations only for a single intensity value equal to the effective intensity in the experiment. 

\subsection{Data analysis}
The photoelectron spectra were background-corrected and filtered w.r.t. fluctuations in FEL pulse energy and photon energy. 
The effective intensity $I_\mathrm{eff}$ was calibrated from the AT splitting taken at the maximum pulse energy according to 
\begin{equation}\label{Eq.Ieff}
    I_\mathrm{eff} = 0.5 \epsilon_0 c \left(\frac{\hbar \Omega}{\mu}\right)^2  \, .
\end{equation}
To this end the AT splitting $\hbar \Omega$ was deduced by fitting the corresponding photoelectron spectrum with a sum of three Gaussian functions. 
For all other pulse energies, the prediction by Eq.\,\ref{Eq.Ieff} was plotted as dashed lines in Fig.\,\ref{fig2}a. 
The Rabi period was calculated based on the determined effective FEL intensity. 

The low intensity contribution in the data shown in Fig.\,\ref{fig3}a was removed by fitting the data with a sum of three Gaussians of which only the amplitude was fitted as free parameter. 
The fitted Gaussian in the center was subtracted from the data. 
For the data shown in Fig.\,\ref{fig4}a, the low intensity distribution was removed by subtracting the photoelectron spectrum shown in Fig.\,\ref{fig2}c scaled in amplitude to account for the different pulse energies used in the two data sets. 

To determine the ratio between the upper and lower photoelectron band shown in Fig.\,\ref{fig4}c, we computed the integral of photoelectron intensity in the upper/lower band for a photon energy of 21.22\,eV (at $1s^2 \rightarrow 1s2p$ resonance) and divided the values. 

\section*{Acknowledgements}
The following funding is acknowledged: Bundesministerium f\"ur Bildung und Forschung (BMBF) \textit{LoKo-FEL} (05K16VFB) and \textit{STAR} (05K19VF3), European Research Council (ERC) Starting Grant \textit{MULTIPLEX} (101078689), Deutsche Forschungsgemeinschaft (DFG) RTG 2717 and grant 429805582 (project SA 3470/4-1) and project STI 125/24-1, Baden-W\"urttemberg Stiftung Eliteprogram for Postdocs, Swedish Research Council and Knut and Alice Wallenberg Foundation, Sweden, Danish Agency for Science, Technology, and Innovation for funding through the instrument center DanScatt. 
The research leading to this result has been supported by the COST Action CA21101 “Confined Molecular Systems: From a New Generation of Materials to the Stars (COSY).

\section*{Author contributions}
L.B. conceived the experiment with input from U.S. and M.W.. 
E.A., M.D., A.D., I.N., F.P. and P.S. implemented and characterized the spectral phase shaping of the FEL. 
E.A., A.B., L.G., Mi.M., G.P., A.S., M.T. and M.Z. optimized the machine and the laser beam parameters. 
C.C, M.D.F. and O.P. managed the end-station. 
F.R., B.A., G.C., K.D., S.D.G., N.G., S.H., F.L., Y.L., C.M., Mo.M., A.M., Ma.M., A.N., N.P., K.C.P., N.R., F.S., D.U., B.W., C.C, M.D.F., O.P. and L.B. performed the experiment with input from U.S., M.W., R.F., B.v.I., T.L., G.S. and R.J.S.. 
F.R. analyzed the data under the supervision of L.B.. 
U.S. provided the theoretical calculations. 
L.B. wrote the manuscript with input from all authors. 

\section*{Disclosures}
The authors declare no conflicts of interest.

\section*{Data availability}
The data that support the findings of the study are openly available.

\section*{Code availability}
The code that supports the findings of the study is available from the corresponding authors on reasonable request.

\bibliography{literature/arxiv_natbib}
\end{document}


\title{Supplementary Information: Strong-field quantum control in the extreme ultraviolet using pulse shaping}

\author{Fabian Richter}
\affiliation{Institute of Physics, University of Freiburg, Hermann-Herder-Str. 3, D-79104 Freiburg, Germany}
\author{Ulf Saalmann}
\affiliation{Max-Planck-Institut für Physik komplexer Systeme, N\"othnitzer Str. 38, 01187 Dresden,
Germany}
\author{Enrico Allaria}
\affiliation{Elettra-Sincrotrone Trieste S.C.p.A., 34149 Basovizza (Trieste), Italy}
\author{Matthias Wollenhaupt}
\affiliation{Institute of Physics, University of Oldenburg, Carl-Von-Ossietzky-Str. 9-11, 26129
Oldenburg, Germany}
\author{Benedetto Ardini}
\affiliation{IFN-CNR, Dipartimento di Fisica, Piazza L. Da Vinci 32, 20133 Milan, Italy}
\author{Alexander Brynes}
\author{Carlo Callegari}
\affiliation{Elettra-Sincrotrone Trieste S.C.p.A., 34149 Basovizza (Trieste), Italy}
\author{Giulio Cerullo}
\affiliation{IFN-CNR, Dipartimento di Fisica, Piazza L. Da Vinci 32, 20133 Milan, Italy}
\author{Miltcho Danailov}
\author{Alexander Demidovich}
\affiliation{Elettra-Sincrotrone Trieste S.C.p.A., 34149 Basovizza (Trieste), Italy}
\author{Katrin Dulitz}
\affiliation{Institut für Ionenphysik und Angewandte Physik, Universit\"at Innsbruck, 6020 Innsbruck, Austria}
\author{Raimund Feifel}
\affiliation{Department of Physics, University of Gothenburg, Origov\"agen 6 B, SE-412 96 Gothenburg, Sweden}
\author{Michele Di Fraia}
\affiliation{Elettra-Sincrotrone Trieste S.C.p.A., 34149 Basovizza (Trieste), Italy}
\affiliation{Istituto Officina dei Materiali - CNR (CNR-IOM), Strada Statale 14 – km 163.5, Trieste, 34149, Italy}
\author{Sarang Dev Ganeshamandiram}
\affiliation{Institute of Physics, University of Freiburg, Hermann-Herder-Str. 3, D-79104 Freiburg, Germany}
\author{Luca Giannessi}
\affiliation{Elettra-Sincrotrone Trieste S.C.p.A., 34149 Basovizza (Trieste), Italy}
\affiliation{Istituto Nazionale di Fisica Nucleare - Laboratori Nazionali di Frascati, Via E. Fermi 40, 00044 Frascati, Roma}
\author{Nicolai G\"olz}
\author{Sebastian Hartweg}
\author{Bernd von Issendorff}
\affiliation{Institute of Physics, University of Freiburg, Hermann-Herder-Str. 3, D-79104 Freiburg, Germany}
\author{Tim Laarmann}
\affiliation{Deutsches Elektronen-Synchrotron DESY, Notkestr. 85, 22607
Hamburg, Germany}
\affiliation{The Hamburg Centre for Ultrafast Imaging CUI, Luruper Chaussee 149, 22761 Hamburg, Germany}
\author{Friedemann Landmesser}
\author{Yilin Li}
\affiliation{Institute of Physics, University of Freiburg, Hermann-Herder-Str. 3, D-79104 Freiburg, Germany}
\author{Michele Manfredda}
\affiliation{Elettra-Sincrotrone Trieste S.C.p.A., 34149 Basovizza (Trieste), Italy}
\author{Cristian Manzoni}
\affiliation{IFN-CNR, Dipartimento di Fisica, Piazza L. Da Vinci 32, 20133 Milan, Italy}
\author{Moritz Michelbach}
\author{Arne Morlok}
\affiliation{Institute of Physics, University of Freiburg, Hermann-Herder-Str. 3, D-79104 Freiburg, Germany}
\author{Marcel Mudrich}
\affiliation{Department of Physics and
Astronomy, Aarhus University, Ny Munkegade 120, DK-8000 Aarhus, Denmark}
\author{Aaron Ngai}
\affiliation{Institute of Physics, University of Freiburg, Hermann-Herder-Str. 3, D-79104 Freiburg, Germany}
\author{Ivaylo Nikolov}
\author{Nitish Pal}
\affiliation{Elettra-Sincrotrone Trieste S.C.p.A., 34149 Basovizza (Trieste), Italy}
\author{Fabian Pannek}
\affiliation{Institute for Experimental Physics, University of Hamburg, Luruper Chaussee 149, 22761
Hamburg, Germany}
\author{Giuseppe Penco}
\author{Oksana Plekan}
\author{Kevin Charles Prince}
\affiliation{Elettra-Sincrotrone Trieste S.C.p.A., 34149 Basovizza (Trieste), Italy}
\author{Giuseppe Sansone}
\affiliation{Institute of Physics, University of Freiburg, Hermann-Herder-Str. 3, D-79104 Freiburg, Germany}
\author{Alberto Simoncig}
\affiliation{Elettra-Sincrotrone Trieste S.C.p.A., 34149 Basovizza (Trieste), Italy}
\author{Frank Stienkemeier}
\affiliation{Institute of Physics, University of Freiburg, Hermann-Herder-Str. 3, D-79104 Freiburg, Germany}
\author{Richard James Squibb}
\affiliation{Department of Physics, University of Gothenburg, Origov\"agen 6 B, SE-412 96 Gothenburg, Sweden}
\author{Peter Susnjar}
\author{Mauro Trovo}
\affiliation{Elettra-Sincrotrone Trieste S.C.p.A., 34149 Basovizza (Trieste), Italy}
\author{Daniel Uhl}
\affiliation{Institute of Physics, University of Freiburg, Hermann-Herder-Str. 3, D-79104 Freiburg, Germany}
\author{Brendan Wouterlood}
\affiliation{Institute of Physics, University of Freiburg, Hermann-Herder-Str. 3, D-79104 Freiburg, Germany}
\author{Marco Zangrando}
\affiliation{Elettra-Sincrotrone Trieste S.C.p.A., 34149 Basovizza (Trieste), Italy}
\author{Lukas Bruder}
\email{lukas.bruder@physik.uni-freiburg.de}
\affiliation{Institute of Physics, University of Freiburg, Hermann-Herder-Str. 3, D-79104 Freiburg, Germany}

\maketitle

\section{Spatial intensity distribution in the interaction volume}
\begin{figure}[b!]
\centering\includegraphics[width=0.8\linewidth]{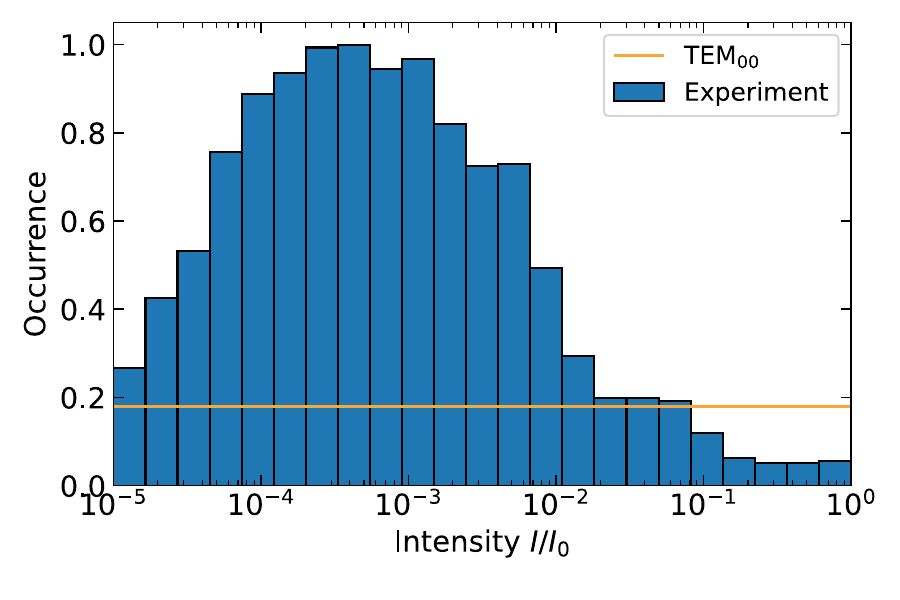}
\caption{Histogram of the intensities in the ionization volume. Blue: measured distribution, orange: theoretical distribution for TEM$_{00}$ mode. Intensities are given relative to the peak intensity $I_0$.
}
\label{SIfig2}
\end{figure}

The spatial intensity distribution in the ionization volume of the experiment was measured with a Hartmann wavefront sensor. 
The atomic jet target has a width of $\approx 0.2$\,mm along the FEL propagation direction, thus the intensity in the direction of propagation can be assumed to be constant. 
To visualize the intensity distribution in the transverse mode, we generated a histogram of the intensity values measured in the ionization volume (Fig.\,\ref{SIfig2}). 
The experimental distribution (blue) is compared to an ideal Gaussian TEM$_{00}$ mode (orange). 
While the TEM$_{00}$ mode is characterized by an equal relative occurrence of all intensity values in the ionization volume, the actual intensity occurrences measured in the experiment show a maximum at intensities roughly three orders of magnitude lower than the peak intensity $I_0$. 
Hence, in the experiment a much larger fraction of He atoms in the ionization volume were excited by low intensities than expected theoretically. 
At these low intensities, the AT splitting is too small to be resolved. 
This rationalizes the appearance of a pronounced peak in the center of the measured photoelectron spectra not showing an AT splitting. 

\section{Reproducibility}
\begin{figure}[h!]
\centering\includegraphics[width=0.8\linewidth]{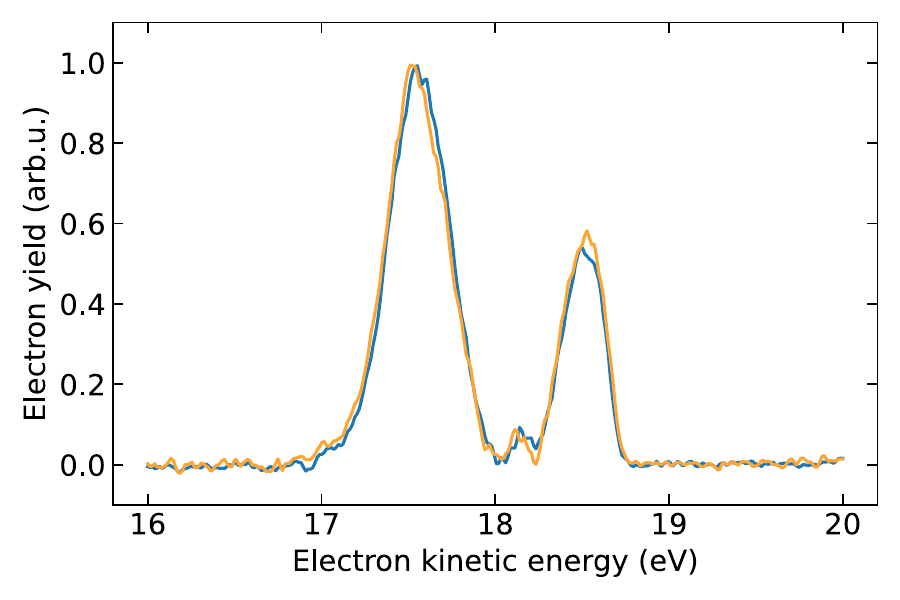}
\caption{Photoelectron spectra for GDD=135\,fs$^2$ of the XUV pulses. The low intensity contribution has been subtracted as done for Fig.\,3a in the main text. 
Orange/blue was taken before/after acquiring the data set in Fig.\,3a (main text).}
\label{SIfig1}
\end{figure}

Fig.\,\ref{SIfig1} shows examples of raw photoelectron spectra for GDD\,= 135\,fs$^2$ taken before and after acquiring the data shown in Fig.\,3a of the main text. 
Very good agreement between the two spectra is found, even though the chirp settings of the seed laser and thus of the FEL were changed in the range -1127\,fs$^2$ to +695\,fs$^2$ over the course of several hours between the two measurements. 
This shows the high reproducibility of the XUV pulse shaping method implemented here. 

\section{Analysis of the adiabaticity}
\begin{figure}[h!]
\centering\includegraphics[width=\linewidth]{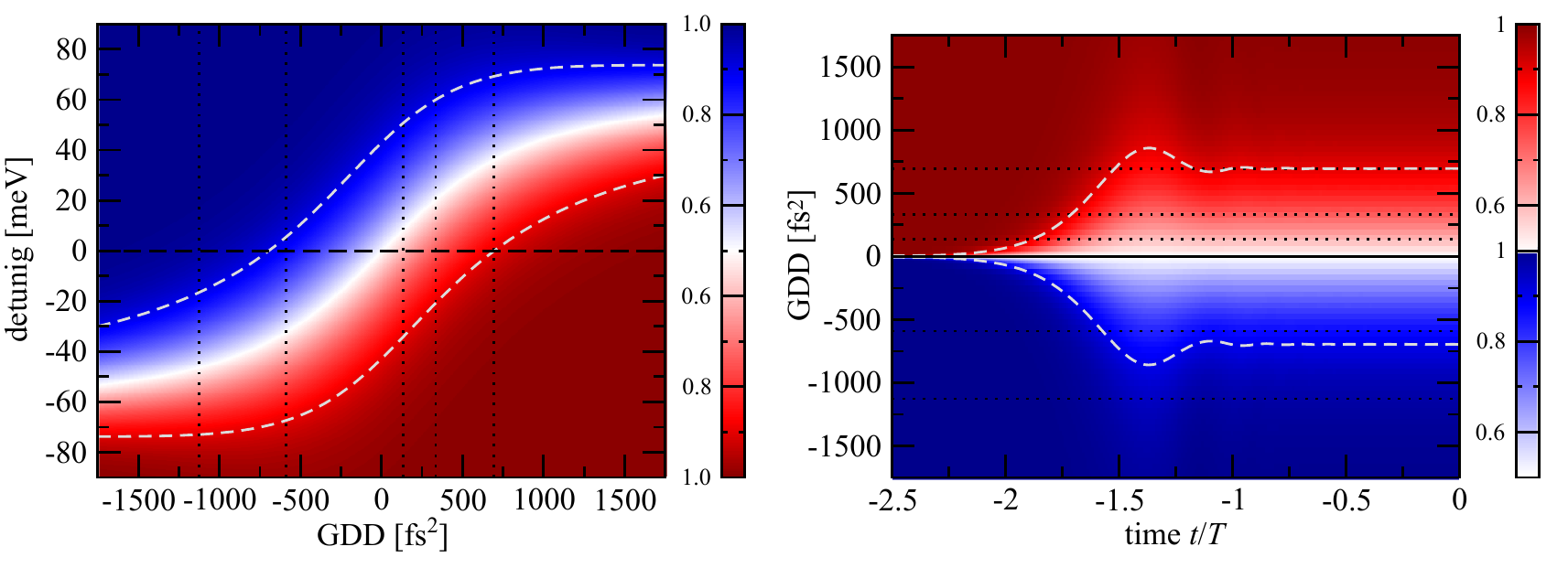}
\caption{Occupation of the AT states during the light-matter interation. 
Left: Dominant occupation of either the upper (blue) or the lower (red) AT state at the peak of the FEL pulse ($t=0$) as a function of the GDD and the detuning from the resonance. The white dashed lines mark the boundary where the corresponding state is occupied with 90\,\% probability. 
For resonant driving (horizontal dashed line) the occupation exceeds 90\,\% only for one GDD value from the experiment, that is -1127\,fs$^2$. 
Right: Occupation probability of the initially occupied AT state  as a function of time $t$ (in units of the pulse duration $T$) for resonant driving. Since for negative and positive GDDs the initially occupied states are the upper and lower AT state, respectively, different colors are used. 
If the occupation remains $\gtrsim 90$\,\% until the peak of the pulse ($t=0$), which is the case for $|\mathrm{GDD}| \gtrsim750\mathrm{fs}^2$, the dynamic is adiabatic. As in the left panel, the dashed lines mark occupation of the initial state with 90\,\% probability. 
In both panels the 5 experimental GDDs are marked with dotted lines. All calculations are done for a driven two-level system with the energy levels and the dipole coupling of Helium for $T=49.3$\,fs and $I=6{\times}10^{14}$W/cm$^2$. 
}
\label{SIfig4}
\end{figure}
In Fig.\,\ref{SIfig4}, we calculated the occupation probability of the upper/lower AT state correlating to the $1s2p$ bare state for our experimental parameters using two-level model. 
The analysis reveals that the population transfer is only adiabatic for a frequency chirp with values of $|\mathrm{GDD}|\gtrsim750\mathrm{fs}^2$. 
Hence, the majority of our quantum control experiment is conducted in the non-adiabatic regime. 
However, the analysis also shows, that the conditions for rapid adiabatic passage can be generally reached with our approach.

\section{Influence of two-photon ionization via nearby states}
\begin{figure}[h!]
\centering\includegraphics[width=\linewidth]{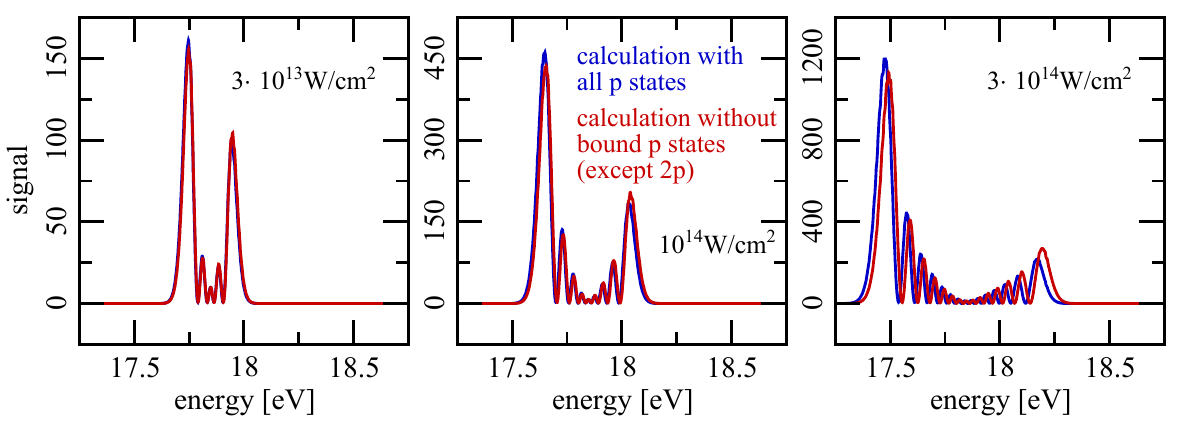}
\caption{Calculated photoelectron spectra for resonant driving (pulse duration: 52.6\,fs, GDD = 0\,fs$^2$) for three different peak intensities (as labeled). Blue: Including all relevant He states. Red: suppressing two-photon ionization pathways via near-resonant states except for the 1s2p state.}
\label{SIfig5}
\end{figure}
In Fig.\,\ref{SIfig5} we investigate the possible contribution of interfering two-photon ionization pathways as a reason for the observed asymmetry in the photoelectron spectra. 
We compare the photoelectron spectrua obtained for the full TDSE-SAE model (blue) with a modified model, where photoionization via any bound state except for the $1s2p$ state is suppressed (red). 
The latter case eliminates interference of multiple photoionization paths. 
The strong similarity between both photoelectron spectra confirms, that photoionization paths via states energetically close to the $1s2p$ state play a negligible role. 
In particular, both spectra exhibit a clear asymmetry between the upper/lower AT states for a range of FEL intensities. 
From this we conclude that the observed asymmetry effect is not due to the interference with two-photon ionization pathways.